\documentclass[conference]{IEEEtran}
\IEEEoverridecommandlockouts
\usepackage{cite}
\usepackage{amsmath,amssymb,amsfonts}
\usepackage{graphicx}
\usepackage{textcomp}
\usepackage{xcolor}
\begin{document}

\title{Concept of Time-Reversal Characteristic Modes in Non-Free-Space Environments}

\author{
\IEEEauthorblockN{1\textsuperscript{st} Chenbo Shi}
\IEEEauthorblockA{
\textit{School of Electronic Science and Engineering} \\
\textit{University of Electronic Science and Technology of China} \\
Chengdu, China \\
chenbo\_shi@163.com}
\and
\IEEEauthorblockN{2\textsuperscript{nd} Jin Pan}
\IEEEauthorblockA{
\textit{School of Electronic Science and Engineering} \\
\textit{University of Electronic Science and Technology of China} \\
Chengdu, China \\
panjin@uestc.edu.cn}
}

\maketitle

\begin{abstract}
Characteristic modes possess a natural environmental interpretation in the current domain because the surrounding scene is carried by the Green function used to construct the impedance operator. An equally direct physical interpretation is less evident in the scattering domain once the background itself participates in propagation and feedback. This paper introduces a field-level definition of time-reversal characteristic modes based on the \emph{differential scattered field} relative to a prescribed background. A characteristic state is identified when this additional field, after time reversal and propagation through the same background, regenerates the same differential scattering state up to a scalar modal factor. The concept is verified in two distinct non-free-space settings. For a finite structural background, it reproduces established substructure characteristic modes and agrees with a common-origin spherical-wave realization. For a PEC half-space, it agrees with the modes obtained from the half-space Green-function impedance operator, although the infinite plane has no natural finite-object transition matrix. The results show that the physical modal statement can remain unchanged even when the numerical representation of the environment is fundamentally different.
\end{abstract}

\begin{IEEEkeywords}
characteristic modes, time reversal, differential scattering, scattering dyadic, substructure, half-space.
\end{IEEEkeywords}

\section{Introduction}
Characteristic mode theory (CMT) identifies intrinsic electromagnetic states without prescribing a particular excitation \cite{ref_CM_Harrington}. A central lesson of unified characteristic-mode formulations is that the surrounding environment enters through the operator that describes field propagation and feedback \cite{ref_Unified_I}. In the current domain this dependence has a direct physical meaning: the impedance operator is constructed from the Green function appropriate to the declared electromagnetic background. Changing the scene changes that Green function, and hence the numerical impedance operator, while the physical question remains the same---how does a current-generated field propagate through its environment and react back on the current?

The corresponding scattering-domain picture is less natural because no single exterior-wave representation remains equally convenient as the background changes. For a bounded object in free space, incoming and outgoing spherical waves provide a natural basis and the transition matrix compactly describes their coupling. With a finite structural background, however, the relevant response is the scattering added relative to that background, so the background scattering process enters explicitly. For an extended interface, the background is part of the propagation scene itself and cannot be treated as another bounded object with a conventional finite-object transition matrix. Thus the representation changes markedly even though the underlying field-feedback question is the same.

We formulate that question directly at the field level. A background is declared first, and the modal object is the scattered field added when the complete configuration replaces that background. Time reversal is then applied to this \emph{differential scattered field} while the same background remains in place. A characteristic state is one for which the returned field regenerates the same additional scattering state up to a scalar modal factor. This definition attaches the mode to a physical change of the electromagnetic scene, rather than to a particular exterior-wave basis.

Two complementary examples establish the concept. A finite-background problem shows that the field definition recovers conventional substructure characteristic modes and their spherical-wave realization. A PEC half-space then removes the finite-scatterer representation of the background entirely: the current-domain reference is formed from the half-space Green function, while the same differential-field time-reversal statement is used in the scattering domain. Their agreement demonstrates that the characteristic state is carried by the background-referenced field physics, not by the choice of a scene-specific matrix representation.

\section{Physical Definition in the Scattering Domain}
\subsection{Differential scattering is the modal object}
Consider a complete configuration and a prescribed electromagnetic background, both excited by the same incident state. Let their scattered fields be \(\mathbf E_{\mathrm{full}}^{s}\) and \(\mathbf E_{\mathrm b}^{s}\), respectively. We define
\begin{equation}
 \mathbf E_{\Delta}^{s}=\mathbf E_{\mathrm{full}}^{s}-\mathbf E_{\mathrm b}^{s}.
 \label{eq:dfield}
\end{equation}
Equation~\eqref{eq:dfield} is a field statement. The background may be represented analytically or numerically, while the differential field itself always isolates the electromagnetic change introduced by the object or region of interest.

The associated differential scattering dyadic is defined physically by the mapping from an incident plane-wave amplitude to the corresponding outgoing differential far-field amplitude. Let \(\hat{\boldsymbol{k}}'\) denote the incident propagation direction and \(\hat{\boldsymbol{k}}\) the outgoing propagation direction. With \(\mathbf F_{\Delta}\) denoting the physical differential far-field coefficient and \(\mathbf E_0\) the incident-field amplitude, we use
\begin{equation}
 \mathbf F_{\Delta}(\hat{\boldsymbol{k}})
 =-\frac{4\pi}{jk}\,
 \overline{\overline{\mathcal S}}_{\Delta}
 (\hat{\boldsymbol{k}},\hat{\boldsymbol{k}}')\cdot
 \mathbf E_0(\hat{\boldsymbol{k}}').
 \label{eq:sdyadic}
\end{equation}
Here \(4\pi/(jk)\) follows the field convention used throughout this paper. The symbol \(\overline{\overline{\mathcal S}}_{\Delta}\) is reserved for this field-level differential scattering dyadic and is distinct from the spherical-wave scattering matrix introduced later.

\subsection{Time-reversal closure}
Start from an outgoing differential scattered field. Reverse its propagation and phase-conjugate the physical field so that it returns through the \emph{same} electromagnetic background. The background controls how the reversed field propagates, reflects, attenuates, or interacts before reaching the structure again. A characteristic state is reached when this returned field produces a differential response with the same outgoing field pattern. The physical experiment is therefore identical in every scene: the scene changes the propagation law, while the closure condition remains unchanged.

After reciprocal/time-reversal pairing and numerical discretization, the self-reproduction condition can be represented as
\begin{equation}
 \mathbf S_{\Delta}^{\mathrm{TR}}\mathbf f_n^{*}
 =\sigma_n\mathbf f_n,
 \label{eq:trclosure}
\end{equation}
where \(\mathbf f_n\) represents the differential scattering pattern. For reciprocal systems this anti-linear problem can be solved conveniently by a Takagi factorization, but the physical definition is the time-reversal closure itself.

The same statement has a current-domain form,
\begin{equation}
 \mathbf Z\mathbf I_n
 =\sigma_n^{-1}\mathbf R^{*}\mathbf I_n^{*},
 \label{eq:momtr}
\end{equation}
where \(\mathbf Z\) is the impedance operator formed with the Green function of the prescribed background. The current-domain and scattering-domain statements therefore refer to the same background propagation-and-return physics. For a reciprocal lossless background,
\begin{equation}
 \mathbf R^{*}=\mathbf R,
 \qquad
 (\mathbf I_n^{*})^{*}=\mathbf I_n,
 \label{eq:involution}
\end{equation}
so the time-reversal operation is explicitly involutory. In the conventional lossless CMT notation,
\begin{equation}
 \mathbf Z\mathbf I_n
 =(1+j\lambda_n)\mathbf R\mathbf I_n,
 \qquad
 t_n=-\frac{1}{1+j\lambda_n},
 \qquad
 \sigma_n=|t_n|.
 \label{eq:classic}
\end{equation}

\section{Two Realizations of the Same Field Definition}
\subsection{Finite structural background}
For a finite background, spherical waves remain a useful numerical representation of the exterior field. Let \(\boldsymbol{\mathsf T}\) and \(\boldsymbol{\mathsf T}_{\mathrm b}\) denote the transition matrices of the complete and background configurations, computed about the same global expansion origin. We use
\begin{equation}
 \boldsymbol{\mathsf S}^{\mathrm{sph}}
 =\mathbf{1}+2\boldsymbol{\mathsf T},
 \qquad
 \boldsymbol{\mathsf S}_{\mathrm b}^{\mathrm{sph}}
 =\mathbf{1}+2\boldsymbol{\mathsf T}_{\mathrm b},
 \label{eq:sphS}
\end{equation}
with identical normalization, ordering, and truncation. The established finite-background characteristic operator can be written as \cite{ref_myBCM}
\begin{equation}
 \mathbf T_{\mathrm{sub}}=
 \frac{1}{2}\left[
 \boldsymbol{\mathsf S}^{\mathrm{sph}}
 (\boldsymbol{\mathsf S}_{\mathrm b}^{\mathrm{sph}})^{\mathrm H}
 -\mathbf{1}\right].
 \label{eq:tsub}
\end{equation}
For the lossless PEC validation considered here, the same modal-strength spectrum follows from the common-origin differential transition matrix,
\begin{equation}
 (\boldsymbol{\mathsf T}-\boldsymbol{\mathsf T}_{\mathrm b})
 \mathbf g_n^{*}=\sigma_n\mathbf g_n.
 \label{eq:tdiff}
\end{equation}
These spherical-wave equations are a convenient discrete realization of the finite-background scattering problem. The physical definition remains Eq.~\eqref{eq:dfield}: the mode belongs to the scattering change from the background scene to the complete scene.

\subsection{PEC half-space}
The distinction between definition and representation becomes especially clear for a PEC half-space. The conducting plane extends to infinity and is part of the propagation environment. In the current domain its effect is carried directly by the half-space Green function, which determines both radiation and the field returned to the strip. Assigning the infinite plane a conventional finite-object spherical-wave \(\boldsymbol{\mathsf T}_{\mathrm b}\) is unnecessary and physically unnatural.

The scattering-domain definition is nevertheless identical to the finite-background case. The bare half-space is the declared background, the additional field produced by the strip is the differential scattered field, and that field is time reversed through the same half-space environment. Thus the same characteristic-state definition survives even when the spherical-wave background construction is unavailable.

\begin{figure*}[!t]
 \centering
 \includegraphics[width=0.97\textwidth]{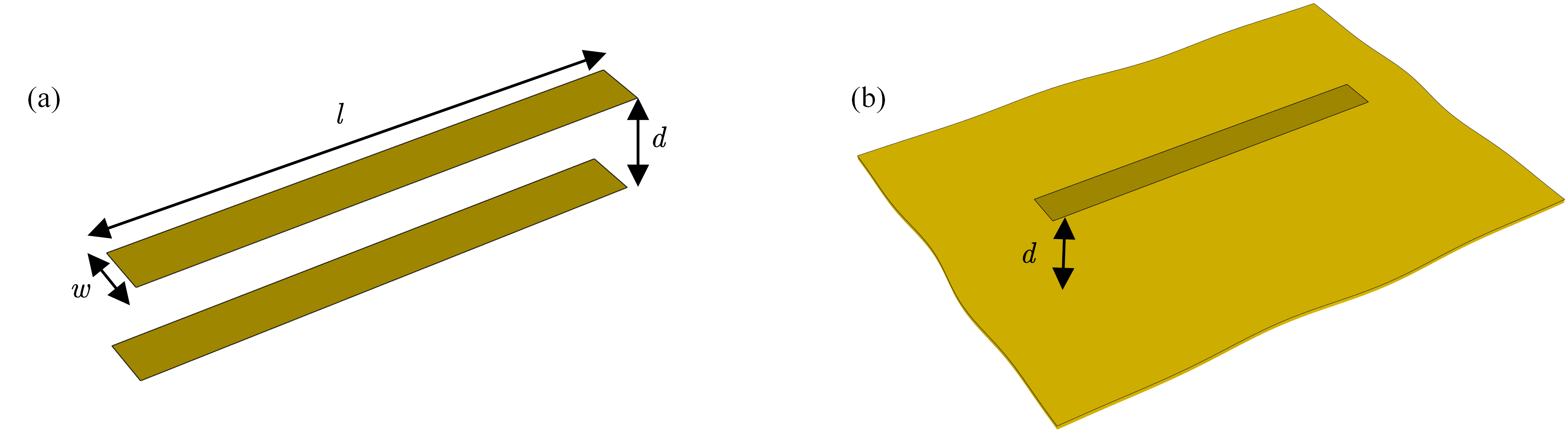}
 \caption{Representative non-free-space environments. Each PEC strip is described by length $l$ and width $w$. (a) A finite structural background, with the upper strip treated as the controlled part and the lower strip retained as the background. (b) A strip above an infinite PEC half-space. The separation from the corresponding background is denoted by $d$.}
 \label{fig:models}
\end{figure*}

\section{Numerical Evidence}
For both examples, \(w=l/10\), and six modal branches are tracked between adjacent electrical-size samples by maximum correlation of their radiated-field patterns \cite{ref_CM_track1}. This enables branch-by-branch comparison through modal crossings.

\subsection{Finite substructure}
\begin{figure}[!t]
 \centering
 \includegraphics[width=3.10in]{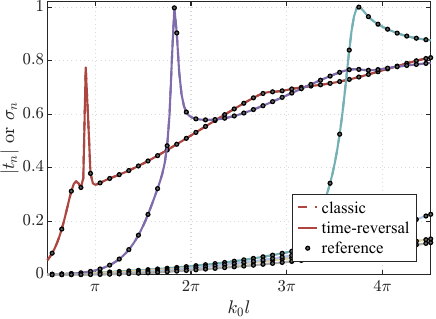}
 \caption{Finite structural background. Dashed lines: classical substructure CMT; solid lines: time-reversal differential-scattering modes; circles: common-origin spherical-wave reference. All six tracked branches are shown.}
 \label{fig:sub}
\end{figure}

For Fig.~\ref{fig:models}(a), the two strips are separated by \(d=l/5\). The spherical-wave realization uses a common global expansion origin and \(l_{\max}=12\). Figure~\ref{fig:sub} compares the established finite-background modes from \eqref{eq:tsub}, the time-reversal modes obtained from the differential scattering dyadic, and the independent common-origin result from \eqref{eq:tdiff}. All six tracked branches are visually coincident.

The agreement has a direct field interpretation. A substructure mode is a characteristic state of the electromagnetic change from the background configuration to the complete configuration. The background determines how the reversed field returns; the differential field determines what scattering belongs to the controlled addition. The conventional substructure operator, the common-origin spherical-wave representation, and the differential scattering dyadic therefore describe the same background-referenced propagation-and-return experiment.

\subsection{PEC half-space}
For Fig.~\ref{fig:models}(b), the strip is separated from the PEC plane by \(d=l/10\). The reference characteristic modes are obtained from \eqref{eq:classic} with the impedance matrix assembled using the PEC half-space Green function. In the scattering domain, the differential field is defined relative to the bare PEC half-space and time reversed through that same environment.

\begin{figure}[!t]
 \centering
 \includegraphics[width=3.10in]{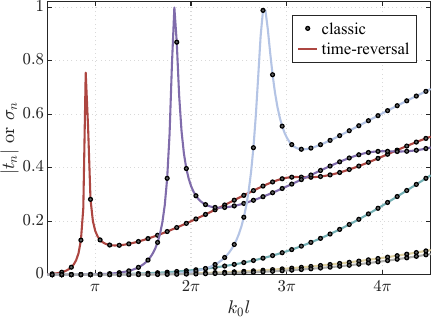}
 \caption{PEC half-space. Circles: characteristic modes obtained from the half-space Green-function impedance operator; solid lines: time-reversal modes obtained from the differential scattering dyadic.}
 \label{fig:half}
\end{figure}

Figure~\ref{fig:half} shows that the six time-reversal branches coincide with the Green-function characteristic modes. This example demonstrates why the field definition is more general than a spherical-wave construction. The background is infinite and has no natural finite-object transition matrix, yet the modal statement is unchanged: the characteristic state is the additional scattering whose time-reversed counterpart returns through the background and regenerates the same additional scattering state.

\section{Scope and Discussion}
The two validations expose the physical distinction that motivates the present definition. In the current domain, environmental dependence is naturally carried by the Green function: once the background is chosen, its Green function specifies how current-generated fields propagate and return. In the scattering domain, an equally natural role is played by the field \emph{change} relative to that background. The finite-substructure and PEC-half-space examples reach the same modal states even though one admits a finite-object spherical-wave realization and the other does not. This indicates that the characteristic state is attached to background-referenced field feedback rather than to any particular scattering basis.

The same concept transfers directly to other electromagnetic scenes without changing its physical definition. In free space, the prescribed background is homogeneous and the usual transition-matrix representation is recovered. For finite structural backgrounds, the differential field measures the scattering added to the background configuration. For extended interfaces and general planar layered media, the environment is carried by the corresponding Green function and the modal object remains the additional scattered field relative to that environment. For periodic structures, the periodic background determines the propagation-and-return law while the same differential-field time-reversal closure defines the characteristic state. Loss does not change this definition: attenuation and absorption modify the returned field and the associated operators, but the modal object remains the scattering change relative to the prescribed lossy background.

This viewpoint separates physical definition from numerical representation. Spherical waves are an efficient discrete basis for bounded exterior scattering, whereas environment-specific Green-function operators are natural when the surrounding scene is embedded directly in propagation. These representations need not be forced into one matrix form. They realize the same physical statement: a characteristic mode is a background-referenced scattering state whose time-reversed field returns through that background and reproduces the same additional scattering state.

\section{Conclusion}
A background-referenced physical definition of time-reversal characteristic modes has been introduced in the scattering domain by taking the differential scattered field as the modal object. The background determines how that field propagates and returns, while the characteristic condition is the self-reproduction of the same additional scattering state under time reversal. For a finite structural background, the definition reproduces established substructure modes and their common-origin spherical-wave realization. For a PEC half-space, it reproduces the characteristic modes obtained from the half-space Green-function impedance operator without introducing a finite-object transition matrix for the infinite plane. Together, the two examples establish the same field-level modal statement across two non-free-space environments with fundamentally different conventional representations.

\end{document}